**High performance THz emitters based on ferromagnetic/nonmagnetic heterostructures**

*Yang Wu, Mehrdad Elyasi, Xuepeng Qiu, Mengji Chen, Yang Liu, Lin Ke,* and *Hyunsoo Yang\**

Dr. Y. Wu, M. Elyasi, Y. Liu, M. Chen, Prof. H. Yang
Department of Electrical and Computer Engineering and NUSNNI-NanoCore, National University of Singapore, 117576 Singapore
E-mail: eleyang@nus.edu.sg

Prof. X. Qiu
Department of Electrical and Computer Engineering, National University of Singapore, 117576 Singapore
Shanghai Key Laboratory of Special Artificial Macrostructure Materials and Technology, Institute of Advanced Study and School of Physics Science and Engineering, Tongji University, Shanghai 200092, China

Dr. L. Ke
Institute of Materials Research and Engineering, A-STAR, 138634 Singapore

[+] Y.W. and M.E. contributed equally to this work.

Terahertz (0.1 – 10 THz) spectroscopy is a powerful characterization method due to its unique properties.[1-3] As the bonding energies of many large molecules are in the range of the THz waves (from 0.5 to 50 meV), THz spectroscopy can be applied to material composition analysis which is important for biological, medical, and chemical applications.[4-6] The propagation properties of THz waves highly depend on the conductivity of materials following the Drude model,[7] so that many materials such as paper, dry woods and semiconductor wafers are transparent to the THz waves.[8] Therefore, THz spectroscopy is utilized as a non-destructive and non-invasive method for conductivity measurements of a wide range of materials.[9-11]

For THz technologies, there is great interest in developing high performance THz sources. For THz time domain spectroscopy (TDS), optical rectification from electro-optical (EO) crystals,[12-15] transient electrical currents in semiconductor antennas,[16] and air plasmas induced by a focused *fs*-laser beam[17-19] are the main methods for THz wave generation. It is noticeable that when EO crystals are excited above their bandgaps, the resonance-enhanced second-order nonlinearity can enhance the THz emission dramatically.[15] THz wave emission has been also observed from other types of devices, such as fs-laser pumped magnetic films,[20-22] nonmagnetic films,[23] and magnetically enhanced semiconductor surfaces.[24] Recently, there have been a few reports showing the potential of nonmagnetic (NM) and ferromagnetic (FM) metallic bilayers as THz sources.[25, 26] However, it has not been proven as useful emitters due to the low intensities.

In this work, we report a THz emitter with excellent performances based on NM/FM heterostructures. The spin currents are first excited by the femtosecond laser beam in the NM/FM heterostructures, and then transient charge currents are generated by the inverse spin Hall effect (ISHE), leading to THz emission out of the structure. The broadband THz waves emitted from our film stacks have a peak intensity exceeding that of 500 μm thick ZnTe crystals (standard THz emitters). Our device is insensitive to the polarization of an incident laser beam which indicates the noise resistive feature. In contrast, the polarization of THz waves is fully controllable by an external magnetic field. We have also fabricated the devices on flexible substrates with a great performance, and demonstrated that the devices can be driven by low power lasers. Together with the low cost and mass productive sputtering growth method for the film stacks, the proposed THz emitters can be readily applied to a wide range of THz equipment. Our study also points towards an alternative approach to characterize spintronic devices with NM/FM bilayers.

The schematic of the device structure is shown in **Figure 1**a. A 800 nm wavelength, 120 femtosecond laser beam is incident on the NM/FM structures along the $z$-axis, and the external magnetic field of 1000 Oe is applied along the $-x$-axis. The THz signal is emitted from the device and its electric field strength is probed by a stroboscopic scheme (see Methods). Figure 1b shows a calculated example of the spin density induced by the laser pulse with respect to time in the NM/FM heterostructures (Supporting Information Section

S1), which provides an overview of spin dynamics in the devices. A large number of spin polarized electrons excited at the interface of NM/FM decay dramatically in the first 0.15 ps and then decrease in a much slower rate ($t > 0.2$ ps), generating spin currents along the *z*-direction, until the system reaches an equilibrium state. The excited diffusive spin currents give rise to fast charge dynamics on a picosecond timescale through ISHE, generating in-plane charge currents which emit an electromagnetic (EM) wave in the THz frequency range. Figure 1c is a typical experimental THz signal with the electric field along the *y*-axis, emitted from the structure of high resistance silicon (HR-Si) substrate/Pt (4 nm)/Co (4 nm)/SiO$_2$ (4 nm).

Heterostructures with different NMs are studied with the FM layer of Co, and the THz signals are shown in **Figure 2**. Figure 2a-d show THz signals with a NM layer of Pt, Ir, Gd, and Ru, respectively, showing a peak at ~2 ps. On the other hand, there is a dip at 2 ps when the NM is Ta or W (Figure 2e, f). Furthermore, no clear peak is observed with the Cu NM layer as shown in Figure 2g. The opposite polarities in the THz signal indicate that the in-plane charge currents have a 180 degree phase shift in their oscillations. The spin currents in the bulk of the NM and the NM/FM interface are converted to charge currents via the inverse spin orbit interaction (SOI), such as ISHE and inverse spin orbital torques (ISOT). Our model indicates that the $E_y$ component of THz emission is dominated by the charge current induced by ISHE (refer to Supporting Information Section S1). Therefore, the sign change in the THz TDS signal can be attributed to the sign of the spin Hall angle ($\theta_{SH}$) of different NMs. Our observations are in good agreement with existing reports that Pt has a positive spin Hall angle, Ta and W have a negative spin Hall angle, and Cu has negligible spin-orbit coupling strength.[27-36] Moreover, our results show Ru, Ir, and Gd possess positive spin Hall angles[30] (see Supporting Information Section S2 for the spin Hall angle measurement of Ru).

Furthermore, negligible THz signal is generated from either Pt or Co alone as shown in Figure 2h, which indicates that a bilayer structure of NM/FM is necessary for effective THz generation (THz emission from a Co film alone is available in Supporting Information Section S4). The reason is that the ISHE induced transient charge current relies on the existence of both a spin source of FM (e.g. Co) and a spin sink of NM with strong SOI (e.g. Pt). Figure 2i shows that there is no THz emission from a FM of CoFeB with perpendicular magnetization anisotropy (PMA) adjacent to a Pt NM layer. Since the magnetization ($\vec{M}$) direction of the CoFeB film is out-of-plane and the net flow of spin currents ($\vec{J}_S$) is also in the same $z$-direction, no charge current ($\vec{J}_C$) is expected to be generated due to ISHE ($\vec{J}_C \propto \theta_{SH} \vec{J}_S \times \vec{M}$). We further extend our experiments to devices with different oxygen levels in the sub/Co (3 nm)/Ta structure by changing the thickness of Ta (Supporting Information Section S3). When the thickness of Ta is 1-3 nm, there is a peak in the THz time domain spectroscopy (TDS) signal, while the opposite signal sign is observed when the thickness of Ta is 5 nm. The reason for this observation is a strong dependence of the oxygen stoichiometry of Ta thin films on its thickness. Indeed, this result is coherent with a recent report that the oxidation of Ta can lead to the change of its effective spin Hall angle.[37]

Optical excitation in a FM/NM bilayer by a femtosecond pulse leads to the demagnetization of a FM layer in the vicinity of the NM layer, which is affected to a large extent through the diffusion of spin currents between the two layers, at different rates for the majority and minority spins.[38, 39] The diffusive spin current in the NM layer with strong SOI gives rise to a bulk charge current through ISHE. In addition, the lack of the inversion symmetry at the interface of a strong SOI material (heavy metal) causes the existence of a Rashba contribution which in its inverse form converts spin to charge currents. This interfacial charge current through ISOTs can be distinguished from the bulk ISHE induced charge current by measuring the THz signal along the $x$-axis. However, due to a much smaller signal intensity, the THz signal along the $x$-axis is not clearly observed in this study. Using our theoretical model (Supporting Information Section S1), we have calculated the ISHE current density (**Figure 3**a), and the resulting electric field of the THz signal along the *y*-axis ($E_y$) (Figure 3a inset). A polarity change in $\theta_{SH}$ changes the sign of $E_y$, as shown in the inset of Figure 3a, which agrees with the experimental results in Figure 2. In order to confirm that the $E_y$ signal is dominated by the ISHE, a thin Cu layer is inserted between the Pt and Co layer, and the measured THz emission signals ($E_y$) are shown in Figure 3b. As the thickness of the Cu layer increases, the intensity of THz signals decreases gradually, indicating the bulk ISHE dominates the THz signal, since an interfacial effect is expected to induce an abrupt change with the Cu layer insertion.

A thickness dependence study has been carried out on sub/Pt/Co/SiO$_2$ and sub/W/Co/SiO$_2$ stacks. As shown in **Figure 4**a, the peak amplitude of THz signals from both samples increases as the thickness of Pt or W increases, and then saturates. Due to the limited spin diffusion in the NM layer, the charge currents are mostly generated in the vicinity of the FM layer where the non-equilibrium spin currents exist. The thickness of Co has been also varied in stacks with Pt and W as NM layers (Figure 4b). The THz peak reaches the maximum value with 1-2 nm Co and then decreases gradually. This phenomenon is correlated to the laser-induced spin diffusion and THz optical absorption in the Co layer. Corresponding theoretical calculation (Supporting Information Section S1) results are in Figure 4c and 4d. Figure 4c shows that the calculated temporal peak of the spin accumulation increases by increasing the thickness of NM ($d_{NM}$), indicating consequent higher THz signals, which is similar to the experimental result in Figure 4a. Figure 4d shows that the temporal peak of the spin accumulation first increases by increasing the thickness of FM ($d_{FM}$) and then decreases, as is also evident in the experimental data (Figure 4b).

From the thickness dependence study, the sample structure has been optimized to sub/ NM (6 nm)/FM (3 nm)/capping layer. It is found that film stacks on glass, sapphire and polyethylene terephthalate (PET) substrates emit much stronger THz waves than the samples on HR-Si wafers (Supporting Information Section S4). Typical THz TDS signals from a 500 µm ZnTe crystal and a glass/W (6 nm)/Co (3 nm)/Al$_2$O$_3$ (3 nm) film stack are compared in **Figure 5**a, and corresponding frequency domain spectra are shown in Fig. 5b. The oscillation features in Figure 5b are due to the micro bubbles in the detection ZnTe crystal. It can be seen that our 12 nm thin films have similar performances in THz generation compared to the 500 µm thick ZnTe crystals. The bandwidth of measured signals is up to 8 THz which is limited by the 120 fs pulse duration of our laser source. A high signal to noise ratio up to 65 dB is achieved in our THz emitters. The comparison of THz emitters between our FM/NM heterostructures and InAs samples is discussed in Supporting Information Section S5.

The polarization dependence of the incident source beam is studied in the Pt NM sample. Unlike a strong linear polarization dependence from a standard THz emitter of <110> ZnTe crystal (Supporting Information Section S6), there is no angular dependence in THz emission from the Pt/Co structure (Figure 5c). Similarly, rotating the helicity of incident light (circular polarization) causes negligible changes in the THz emission (Supporting Information Section S7). The independence of the THz signals on the linear and circular polarization of incident light indicates that the THz generation in the NM/FM stacks does not depend on the non-linear optical response caused by the crystalline structure of the samples, but is mainly attributed to the non-equilibrium spin and charge transport, which is beneficial for stable THz emissions.

In order to study the dependence of the detected THz E-field on the magnetic field direction, the magnetic field of 1000 Oe is rotated in the *xy*-plane for samples with Pt and W NM layers as shown in Figure 5c. The rotation angle ($\theta$) is indicated in Figure 1a. The peaks of THz signals follow a sinusoidal trend for both samples. As the THz emission is dominated by the components perpendicular to the external magnetic field, the projected E-field strength along the *y*-axis should be the strongest when *M* is aligned with the ±*x*-axis ($\theta$ = 90 and 270º), and approach to 0 when *M* is along the ±*y*-axis ($\theta$ = 0 and 180º).

In addition, the Pt/Co stacks are fabricated on flexible PET substrates as shown in the inset of Figure 5d, and THz TDS signals are measured with four different bending curvatures (Supporting Information Section S8). The device performance is not deteriorated with large bending curvatures as shown in Figure 5d. The high performance of our devices on flexible substrates indicates mechanical robustness, which is an advantage over conventional fragile THz emitters and applicable for future wearable devices.

The dependence of THz emission on the incident laser power has been studied for quartz/W/Co/cap samples (Supporting Information Section S9). The peak intensity of TDS signal initially increases with an increase of the laser power, and then shows a saturating trend. A clear THz TDS signal is obtained with a laser energy density down to 0.6 μJ/cm$^2$, which indicates that our THz emitter can be powered by a low power laser source of 0.15 mW (Supporting Information Section S10). Enabling the use of a low power laser source can lower the cost of total THz systems, in which a fs-laser source is the most expensive part. Our FM/NM heterostructures are prepared by a sputtering method, suggesting that a large wafer-scale film can be deposited, and the wafer can be subsequently diced to the required sample area (~ 8 mm diameter in our setup). Such a low cost fabrication methodology is commercially scalable, especially compared to conventional THz emitters (e.g. EO crystals and photoconductive antennas).

We use a femtosecond laser to excite an ultrafast transient spin current and simultaneously THz emission in NM/FM heterostructures. A low cost (sputtering deposited thin films less than 16 nm), intense (peak intensity exceeding that of 500 μm ZnTe), broadband, noise resistive (not sensitive to the incidence polarization), magnetic field controllable, flexible and low power driven (0.15 mW) THz source is reported. Our results pave the route for developing efficient THz sources based on NM/FM heterostructures. In addition, we have shown that the ISHE from NM/FM bilayers can be characterized by ultrafast laser induced THz emission, which is of importance for spintronics devices.

# Experimental Section

*Sample preparation*: The samples were deposited on HR-Si wafers (> 10,000 Ω/cm$^2$), quartz or PET by sputtering. The HR-Si and quartz substrates were cleaned by acetone and isopropyl alcohol, and PET substrates were cleaned by isopropyl alcohol in an ultrasonic bath. The base pressure of the sputter chamber was 3 × 10$^{-9}$ Torr. The sample holder was continuously rotated during the deposition. The samples are capped with a thin Al$_2$O$_3$ or SiO$_2$ layer to prevent oxidization of the FM/NM bilayers.

*Experimental details.* A laser with the full width at the half maximum of 120 fs, center wavelength of 800 nm, and 1 kHz repetition rate was used for this study. The laser beam was split into two for the stroboscopic sampling; the THz generation was excited by 220 μJ power with a beam diameter of 8 mm, while a much weak power (~ 2 μJ) with a beam diameter of 2 mm was used for THz signal detection (both THz emission and detection beams are not focused). The THz radiation was collected by a parabolic mirror and then focused onto a 500 μm thick ZnTe crystal. Due to the Pockels effect, birefringence is generated when the THz wave shines on the crystal, and subsequently the transmitted detection beam experiences a polarization rotation which can be analyzed by a balanced photodetector system. A pair of magnets was mounted on a rotation stage with magnetic fields (1000 Oe) along the −*x*-axis in a dipole configuration. A wire grid THz polarizer is placed after the samples with wires parallel to the magnetic field direction to define the polarization for the THz wave. The THz generation and detection parts are enclosed in a dry environment with a humidity level of 1.5%.

*Supporting Information*
Supporting Information is available online.

*Acknowledgements*
This research is supported by the National Research Foundation (NRF), Prime Minister's Office, Singapore, under its Competitive Research Programme (CRP award no. NRFCRP12-2013-01).

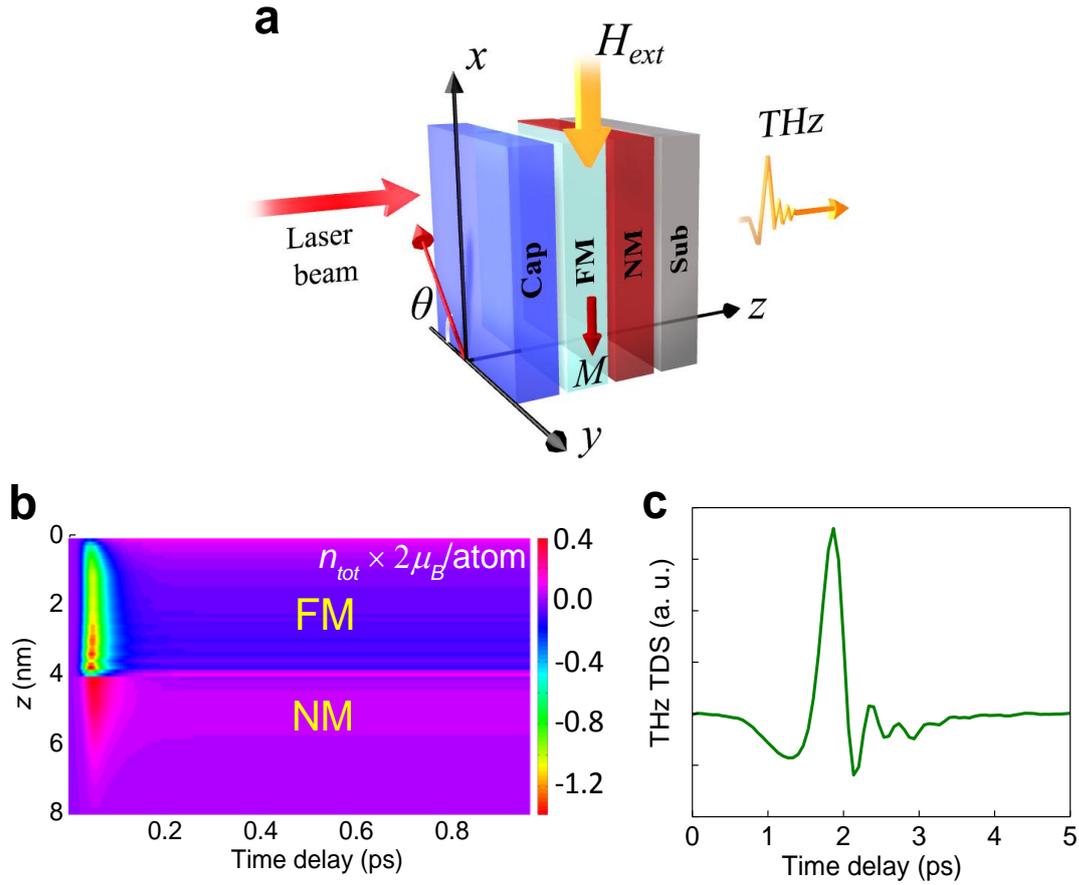

**Figure 1.** a) A typical NM/FM sample structure and the THz emission geometry. The *fs* laser beam is used for pumping the samples. The external magnetic field is along the −*x*-axis. The angle (*θ*) indicates the rotation for either the external magnetic field or the source beam polarization direction. b) Calculated spin dynamic in the FM and NM layers with respect to time. c) A typical experimental THz signal emitted along the *y*-axis induced by transient charge currents due to ISHE.

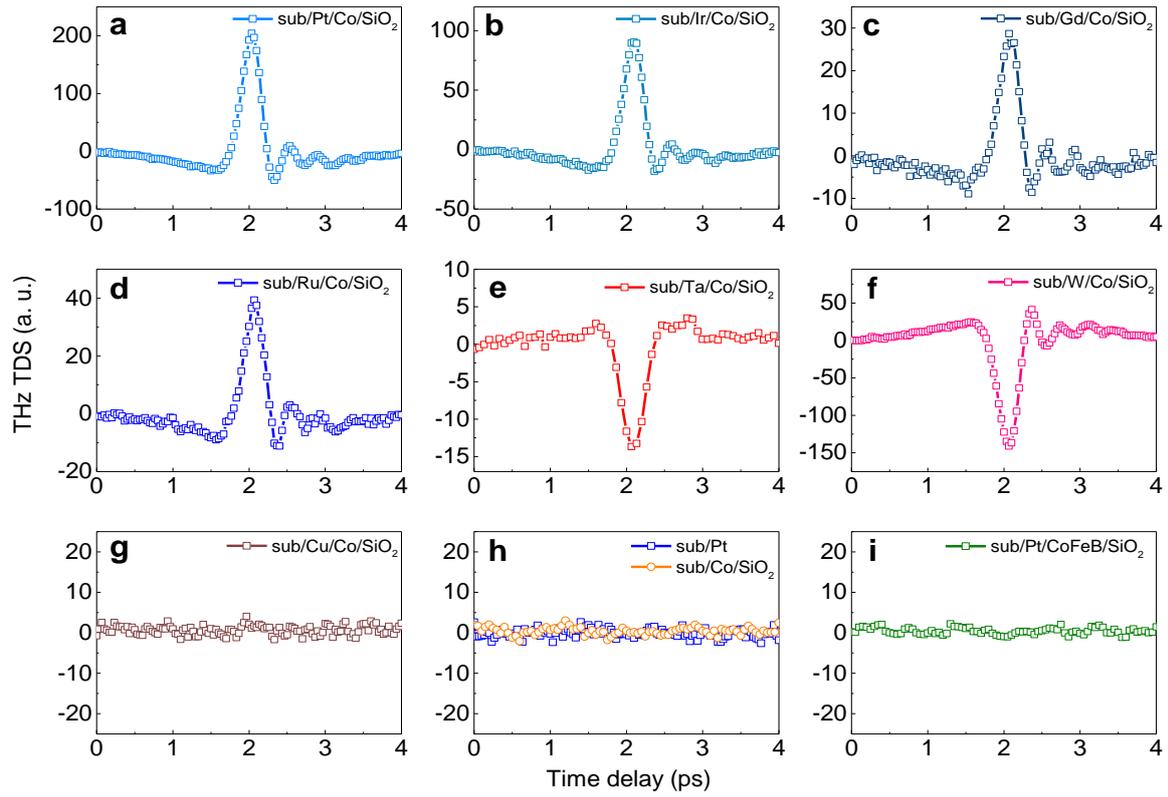

**Figure 2.** THz signals from substrate/NM (4 nm)/Co (4 nm)/SiO$_2$ (4 nm). The NM layer is Pt (a), Ir (b), Gd (c), Ru (d), Ta (e), W (f), and Cu (g). h) Reference samples with Pt (4 nm) or Co (4 nm) layer. i) Pt (4 nm)/Co$_{40}$Fe$_{40}$B$_{20}$ (1 nm)/SiO$_2$ (4 nm) with PMA.

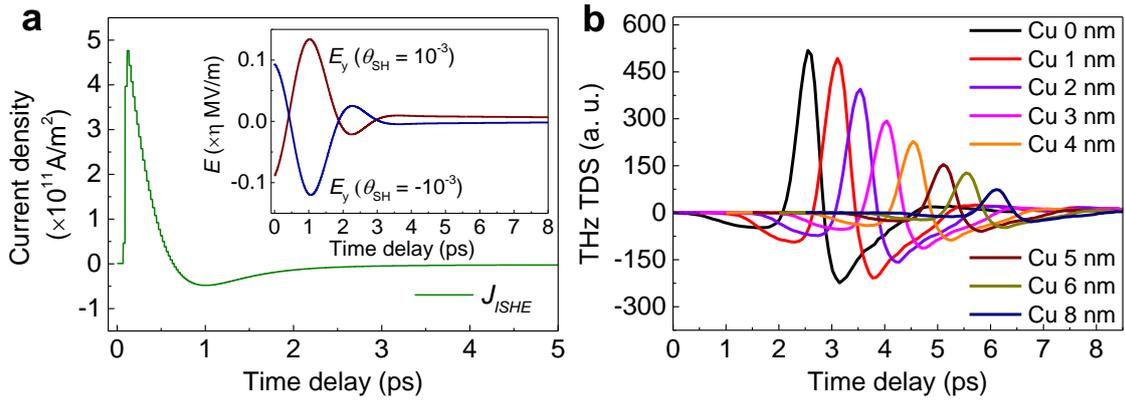

**Figure 3.** a) Calculated $J_{ISHE}$ in the time domain, time "0" is defined as the laser is incident on the sample. The inset is calculated THz emission with different spin Hall angles ($\theta_{SH}$), $10^{-3}$ and $-10^{-3}$. b) THz TDS signal from glass/Pt (6 nm)/Cu (*n* nm)/Co (3 nm)/cap, where *n* = 0, 1, 2, 3, 4, 5, 6, and 8 nm). The data are shifted horizontally for clarity.

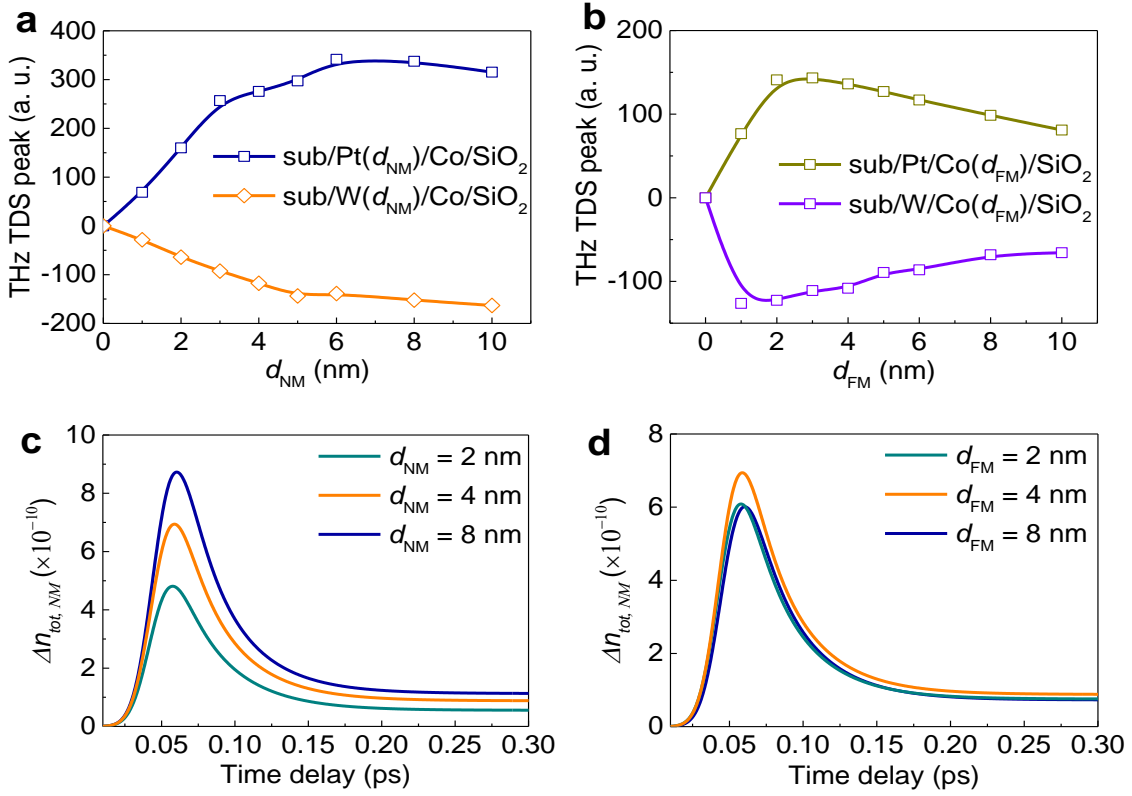

**Figure 4.** a) THz signal by varying the Pt or W thickness in NM/Co (4 nm)/SiO$_2$ (4 nm). b) THz signal by changing Co thickness in NM (4 nm)/Co/SiO$_2$ (4 nm). c) Calculated spin accumulations with 2, 4 and 8 nm NM layer on top of a 4 nm FM layer. d) Calculated spin accumulations with 2, 4 and 8 nm FM with a 4 nm NM layer. The amount of spin accumulation is directly related to the strength of the THz emission. Spin accumulation in NM is $\int_{d_{NM}} \sum_{E} \left( \Delta n_{tot}(E,\uparrow) - \Delta n_{tot}(E,\downarrow) \right)$, where $\Delta n_{tot}(E,\uparrow)$ and $\Delta n_{tot}(E,\downarrow)$ are the total changes in spin-up and spin-down densities, respectively. $d_{NM}$ is the NM thickness, and $d_{FM}$ is the FM thickness.

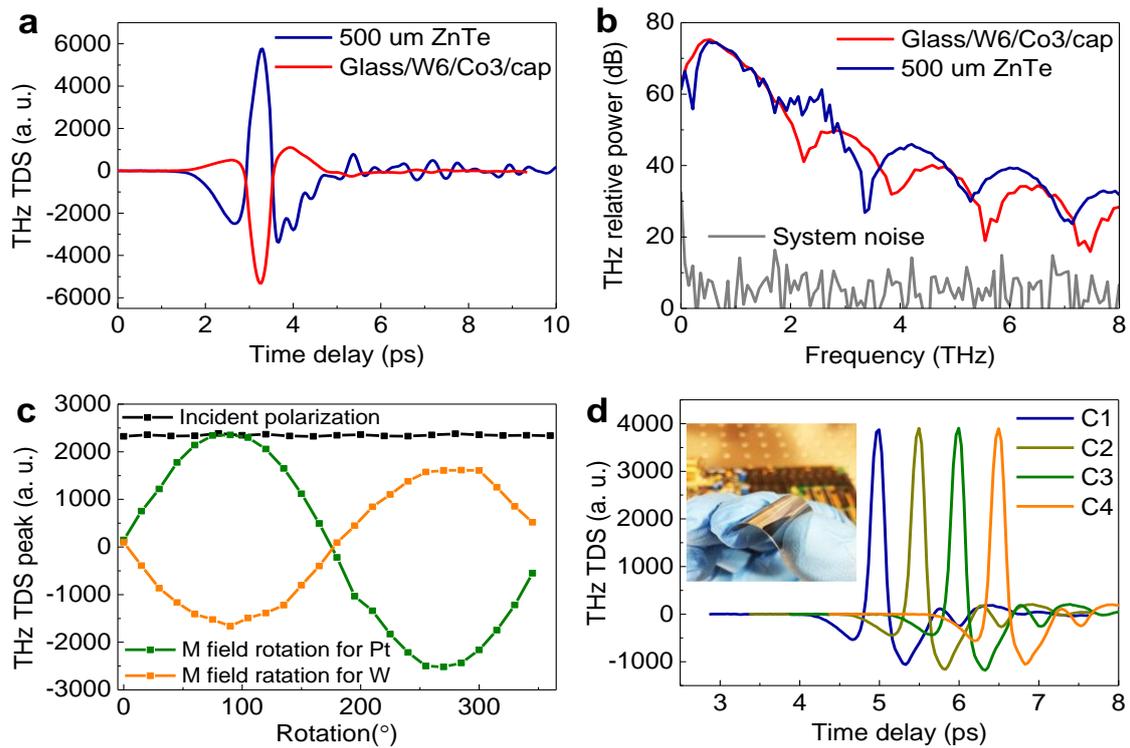

**Figure 5.** a) THz TDS signal acquired from a standard THz generation crystal (500 μm thick <110> cut ZnTe) and our film stacks (glass/6 nm W /3 nm Co /cap). b) Frequency domain data for the THz TDS signals in (a). The system noise level is shown by the gray curve. c) Peak amplitude of THz TDS peak signals with different source beam polarizations for sub/Pt (4 nm)/Co (4 nm)/SiO$_2$ samples (black), and dependence of peak amplitude of THz TDS on the magnetization direction for Pt (olive) and W (orange) NM samples. d) THz TDS signals from samples on flexible PET substrates. The samples are bent to four different levels, C1 ($\kappa$ = 0 m$^{-1}$), C2 ($\kappa$ = 67 m$^{-1}$), C3 ($\kappa$ = 125 m$^{-1}$) and C4 ($\kappa$ = 185 m$^{-1}$). The data are shifted horizontally for clarity. The inset is a photograph of a bent device.